\documentclass[pmlr,twocolumn,10pt]{jmlr} 

\usepackage{algorithm}
\usepackage{algpseudocode}
\usepackage[utf8]{inputenc}
\usepackage{siunitx}

\usepackage[switch]{lineno}

\newcommand{\bx}{\mathbf{x}}
\newcommand{\by}{\mathbf{y}}
\newcommand{\bz}{\mathbf{z}}
\newcommand{\bh}{\mathbf{h}}

\theorembodyfont{\upshape}
\theoremheaderfont{\scshape}
\theorempostheader{:}
\theoremsep{\newline}

\jmlrvolume{LEAVE UNSET}
\jmlryear{2024}
\jmlrsubmitted{LEAVE UNSET}
\jmlrpublished{LEAVE UNSET}
\jmlrworkshop{Conference on Health, Inference, and Learning (CHIL) 2024} 

\title[DDoS]{DDoS: a graph neural network based drug synergy prediction algorithm}

 \author{%
  \Name{Kyriakos Schwarz} \Email{kyriakos.schwarz@uzh.ch}\\
  \Name{Alicia Pliego-Mendieta} \Email{alicia.pliegomendieta@usz.ch}\\
  \Name{Amina Mollaysa} \Email{ maolaaisha.aminanmu@uzh.ch}\\
  \Name{Lara Planas-Paz} \Email{lara.planas-paz@usz.ch}\\
 \Name{Chantal Pauli} \Email{chantal.pauli@uzh.ch}\\
  \Name{Ahmed Allam} \Email{ahmed.allam@uzh.ch}\\
  \Name{Michael Krauthammer} \Email{michael.krauthammer@uzh.ch}\\
  \addr University of Zurich, Switzerland
 }

\linenumbers 

\begin{document}
\nolinenumbers 
\maketitle
\begin{abstract}
Drug synergy arises when the combined impact of two drugs exceeds the sum of their individual effects. While single-drug effects on cell lines are well-documented, the scarcity of data on drug synergy, considering the vast array of potential drug combinations, prompts a growing interest in computational approaches for predicting synergies in untested drug pairs.
We introduce a Graph Neural Network (\textit{GNN}) based model for drug synergy prediction, which utilizes drug chemical structures and cell line gene expression data. We extract data from the largest available drug combination database (DrugComb) and generate multiple synergy scores (commonly used in the literature) to create seven datasets that serve as a reliable benchmark with high confidence. In contrast to conventional models relying on pre-computed chemical features, our GNN-based approach learns task-specific drug representations directly from the graph structure of the drugs, providing superior performance in predicting drug synergies. Our work suggests that learning task-specific drug representations and leveraging a diverse dataset is a promising approach to advancing our understanding of drug-drug interaction and synergy.
\end{abstract}

\paragraph*{Data and Code Availability}
For this study, we used the drug combination dataset from the \textit{DrugComb}~\citep{zheng2021drugcomb} database (version $1.5$) and we retrieved untreated cell line features from cancerrxgene\footnote{\tiny{\url{https://www.cancerrxgene.org/gdsc1000//GDSC1000_WebResources//Data/preprocessed/Cell_line_RMA_proc_basalExp.txt.zip}}}. From the raw data, we constructed new datasets based on the majority voting schemes as described in the paper. The code is publicly available at this \href{https://github.com/uzh-dqbm-cmi/graphnn}{repository}.

\paragraph*{Institutional Review Board (IRB)}
This research did not require IRB approval.

\section{Introduction}
Treatments targeting complex diseases, such as cancer, frequently lead to acquired drug resistance due to patient-specific variability. For instance, drugs targeting only one key component of growth or proliferation pathways may lead to selective pressure and activation of a compensatory mechanism~\citep{Avner2012}, thus making this treatment suboptimal. However, during multi-target inhibition with reduced stringency, drug resistance is less likely. Therefore, the implementation of combination therapy might improve patient treatment as different drugs may target distinct pathways or genes, likely leading to decreased cancer cell survival. In addition to the increased efficacy, combination therapy often reduces toxicity and decreases the likelihood of treatment resistance compared to monotherapy (i.e., single drug) treatments~\citep{Mokhtari2017}.

Due to advancements in high-throughput screening (HTS), the number of drug screening datasets has been growing in recent years (see Appendix ~\ref{sec:database}).  
Meanwhile, a variety of algorithms have been developed for drug synergy prediction. However, these models come with several limitations: 
$1$) Multiple synergy scoring methods have been developed, though each one has its biases and drawbacks. Yet, the majority of studies rely only on one synergy score, i.e., Loewe or Bliss, thus making them biased towards the assumptions of the synergy score they are using~\citep{wooten2021musyc}.
$2$) Most prediction models only utilize a single dataset, which limits the number of drug and cell line combinations a model can learn on. Additionally, this also limits the extent to which the model can be generalized to unknown drugs and unutilized cell lines. 
$3$) Commonly, existing models take advantage of pre-computed chemical properties of drugs. However, with the wider adoption of Graph Neural Networks (\textit{GNN}s) in recent years, it has become more advantageous to learn task-specific representations of drugs in the form of graphs (i.e., chemical structures where atoms are nodes and bonds are edges). 
 
To this end, we propose a Graph Neural Network (\textit{GNN}) based model, \textit{DDoS} (\textbf{D}rug-\textbf{D}rug c\textbf{o}mbination \textbf{S}ynergy model), which offers multiple advantages: 
$1$) It is trained on four different drug synergy scores (Loewe, Bliss, HSA, ZIP), as well as three distinct majority voting-based combinations of those four scores. Hence, it is trained on labels with much higher synergy consensus (or confidence) compared to single score-based labels. 
$2$) It includes samples from \textit{DrugComb}, which is a collection of $34$ distinct drug synergy datasets, in order to learn on a wide variety of drugs and cell lines. $3$) It learns task-specific drug representations, instead of relying on generalized pre-computed chemical features of drugs. 
$4$) It outperforms state-of-the-art baseline models when tested on seven distinct benchmark datasets.

\section{Literature review}
The availability of larger HTS datasets has enabled model development for improved drug combination synergy predictions. These models utilize genomic information of cell lines and physicochemical properties of drugs, or drug-cell interactions for the prediction of optimal drug combinations. Commonly used models include $\mathrm{Random Forest}$, $\mathrm{Support Vector Machine}$, $\mathrm{Gradient Boosting Machine}$, and $\mathrm{Naive Bayes}$~\citep{Kumar2021}. However, more recently Deep Learning models have shown to be advantageous at managing large amounts of data and learning useful representations that often lead to an improved prediction performance. Most of the deep learning models developed for drug synergy prediction tasks are focused on one question: \emph{What is the ideal data representation for both modalities, drugs, and cell lines?}

Initially, deep learning models focused on using a predefined set of chemical features to represent the drugs. For instance,  \textit{DeepSynergy}~\citep{Preuer2018} applied a feed-forward neural network (\textit{FFNN}) on the vector concatenating the two drug and cell line representations, which outperformed previous traditional machine learning models. The model was trained on the \textit{Merck} dataset.
\textit{ MatchMaker}~\citep{Kuru2020}, on the other hand, proposed to use chemical features of each drug from the \textit{DrugComb} database 
 and concatenated it with cell line gene expression features as input to drug-specific subnetworks (DSNs). The DSNs learn a representation for each drug separately conditioned on cell line gene expressions. 

Furthermore, to improve the cell line representation by incorporating multi-omics data, \textit{AuDNNsynergy}~\citep{zhang2021synergistic}
trained three autoencoders using the gene expression, copy number variations, and genetic mutation data of tumor samples from The Cancer Genome Atlas (TCGA) to generate lower dimensional representations. The physicochemical features of individual drugs and the encoded omics data of individual cancer cell lines are used as the input features to a feed-forward neural network that predicts the synergy score. 
Similar works also include Synpred \citep{preto2022synpred}.

Recently, there are more models reported in the literature aimed at incorporating the graph structure of the drugs instead of using predefined physicochemical properties.
 \cite{wang2022deepdds} proposed DeepDDS which uses the molecular
graphs of two drugs and gene expression profiles of one cancer cell line that was treated by these two drugs. They tested GNNs, GAT, and GCN, to extract features from these drugs. \cite{sun2020structure} proposed graph convolutional networks and
deep sets to learn better discriminative convolutional layers compared to conventional GCN, and achieved permutation invariant prediction while capturing complicated interactions. 
 This work was mainly focused on identifying better graph network architecture that can model drug-drug interaction. Similarly, Deepdrug \citep{yin2022deepdrug} works on identifying the best GNN structure for drug-drug interaction.

\section{Model}
Consider a training dataset $D=\{\bx_i, y_i\}_{i=1}^N$, comprising of N samples, where $\bx_i$ denotes a triplet (Drug A, Drug B, and a Cell line), and $y_i\in \{\textit{synergistic, antagonistic}\}$ represents the class label indicating whether the two drugs exhibit synergy against the specified cell line. The objective is to learn a function $f_{\theta}(\bx_i) \rightarrow y_i$ that can predict, for a given new triplet, whether the given pair of drugs demonstrate synergy against the provided cell line.

\begin{figure*}[tb]
  \centering
    \includegraphics[width=10cm, height=11cm]
    {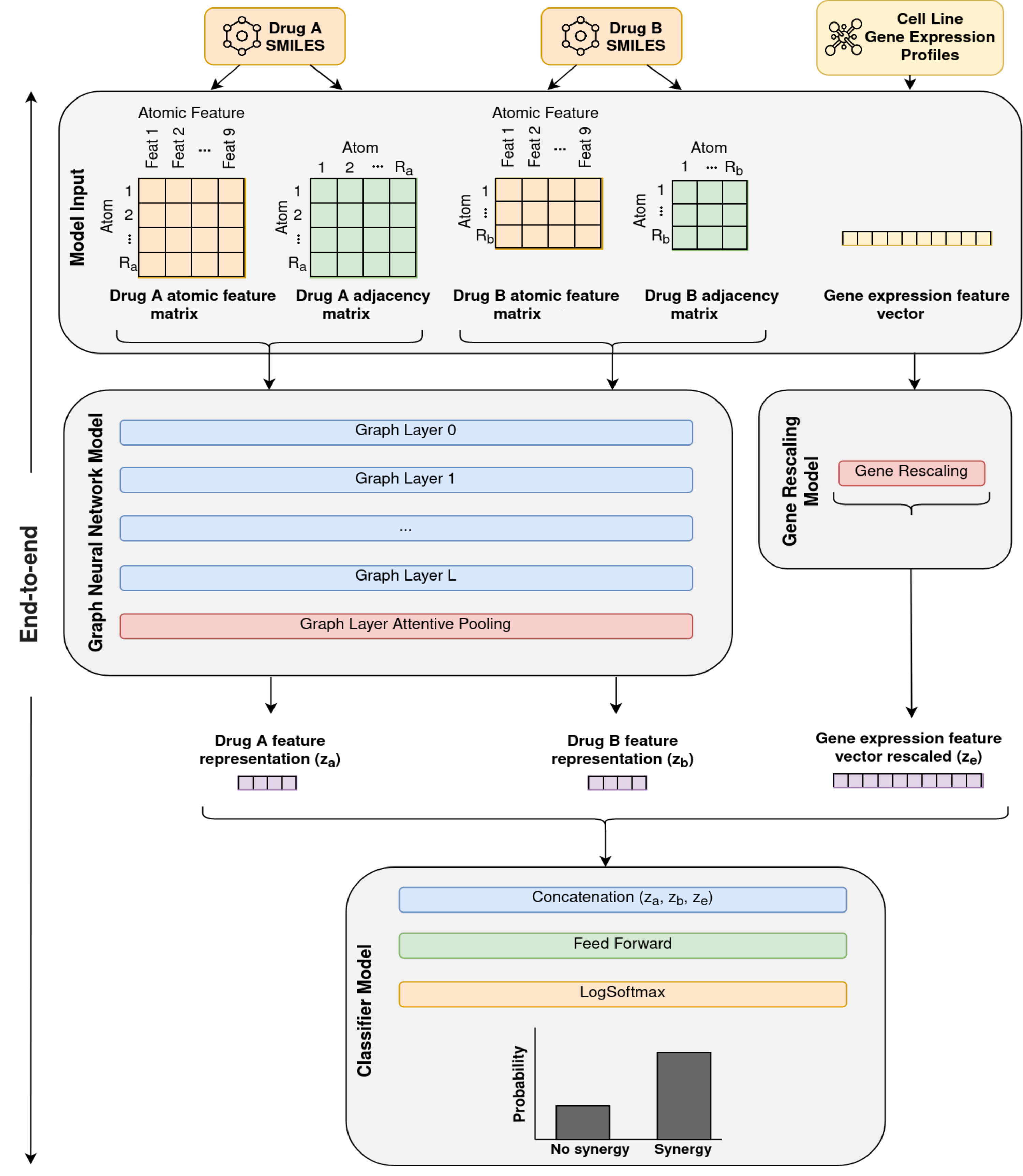}
      \caption{\textbf{\textit{DDoS} model architecture.} 1) The model accepts samples of triplets (\textit{Drug A}, \textit{Drug B}, \textit{Cell Line Gene Expression}) represented by two atomic feature matrices, two adjacency matrices, and one gene expression vector.  2) The atomic and adjacency matrices of each drug are input into a GNN model, producing drug feature representations $\bz_a$ and $\bz_b$ separately. Simultaneously, gene expression features undergo rescaling via the Gene Rescaling model, resulting in a representation vector $\bz_e$. 3) the vectors $\bz_a$, $\bz_b$, and $\bz_e$ are concatenated and fed into a classifier model, determining the probability of synergy for a given sample.
      \label{fig:architecture_overview}}
\end{figure*}

To learn such a function, we first represent the drugs using their molecular graph and apply a graph neural network to map them to their latent representations. Concurrently, we use attention-based feed-forward network to learn a representation for the cell line characterized by its landmark genes. Leveraging a graph neural network, we aim to learn task-specific representations of the drugs, facilitating the accurate prediction of synergistic or antagonistic interactions in the given context. More precisely, our model consists of three main parts (see Fig~\ref{fig:architecture_overview}):

\begin{enumerate}
    \item A Graph Neural Network (\textit{GNN}) model for calculating drug feature representation vectors. 
    \item A \textit{Gene Rescaling} model for rescaling gene expression features, to optimize gene features representation.
    \item A Classifier model (feed-forward neural network followed by a $\mathrm{LogSoftmax}$ function) to compute a probability distribution over the two classes using the learned representation vectors of the \textit{GNN} and \textit{Gene Rescaling} models.
\end{enumerate}

\subsection{Graph Neural Network model}
To better capture the drug-drug interaction, 
we represent drugs as graphs based on their chemical structures. 
A drug with $N_a$ atoms is represented by a graph $\mathcal G = (\mathcal V, \mathcal E)$, $N_a=|\mathcal V|$, where $\mathcal V$ represents the set of nodes (atoms) and $\mathcal E$ as the set of edges (bonds). We then construct a node matrix $ H\in \mathbb R^{N_a\times d}$ where 
$d$ is the number of features used to describe each atom, 
and an adjacency matrix $A\in \mathbb R^{N_a\times N_a}$ where $a_{ij}\in A$ is one if there is a bond between atom $i$ and atom $j$, and zero otherwise.

The graph neural network consists of multiple layers which signifies the ``depth" of the network. Particularly, a layer $l \in \{1, \dots, L\}$ can be viewed as the $l$-hop neighborhood of a node (atom), i.e., the subgraph of all atoms reachable within $l$ steps. Every node, i.e., atom, $ v_i \in \mathcal V$ is initially represented by a node feature vector $\bh_i^{l} \in \mathbb{R}^d$ at layer $l$, and its neighbors are defined by $N_i = \{v_j \in \mathcal V \mid a_{ij}=1\}$. The representation of each node $\bh_i^l$ is updated at layer $l+1$ by aggregating the representations of its neighbors: 

\begin{eqnarray}\label{eq:gnn_general}
\bh_i^{l+1} = \phi(\xi(\{\bh_j^{l} \mid v_j \in N_i\})),
\end{eqnarray}

where $\phi, \xi$ are differentiable functions and $\xi$ is permutation invariant (i.e., generally order invariant). Specifically, we utilize the $\mathrm{GATv2}$ operator~\citep{brody2021attentive} for updating each $\bh_i^l$ (see Eq~\ref{eq:gatv2_update}). This operator provides two advantages, compared to the more commonly used implementation of $\mathrm{GAT}$~\citep{velickovic2017graph}: $1$) the aggregation function $\xi$ is based on the weighted average of neighbors learned representations using attention instead of treating all neighbors with equal importance, and $2$) every node can attend to any other node with dynamic attention while the original $\mathrm{GAT}$ is limited to static attention.

\begin{eqnarray}\label{eq:gatv2_update}
\bh_i^{l+1} = \phi(\alpha_{i,i}\mathbf{W}\mathbf{h}_{i}^{l} +
\sum_{j \in \mathcal{N}_i} \alpha_{i,j}\mathbf{W}\mathbf{h}_{j}^{l}),
\end{eqnarray}

where the attention coefficients $\alpha_{i,j}$ are computed as:

\begin{eqnarray}\label{eq:gatv2_coeff}
\alpha_{i,j} =
\frac{
\exp\left(\mathbf{a}^{\top}\mathrm{LeakyReLU}\left(
[\mathbf{W}\mathbf{h}_{i}^{l} \, \Vert \, \mathbf{W}\mathbf{h}_{j}^{l}]
\right)\right)}
{\sum_{k \in \mathcal{N}_i \cup \{ i \}}
\exp\left(\mathbf{a}^{\top}\mathrm{LeakyReLU}\left(
[\mathbf{W}\mathbf{h}_{i}^{l} \, \Vert \, \mathbf{W}\mathbf{h}_{k}^{l}]
\right)\right)},
\end{eqnarray}

$\mathbf{a}\in \mathbf R^{2d'}, \mathbf{W}\in \mathbf R^{d'\times d}$  are learned parameters, $\phi$ is the $ELU$ activation function and $||$ denotes vector concatenation. 

Subsequently, layer-specific graph representations $\mathbf g^{l}$ are generated by aggregating (i.e., global mean pooling) the updated node representations $\bh_i^{l}$ at layer $l$:

\begin{eqnarray}\label{eq:graph_layer_pooling}
    \mathbf g^{l} = \frac{1}{N_a} \sum_{i=1}^{N_a} \bh_i^{l}, \:\:\: \forall \:\: l\in\{1,2, \dots, L\}
\end{eqnarray}

where the number of \textit{GNN} layers $L$ is a hyperparameter. The final representation of the graph, $\bz$, is obtained as the weighted sum of the layer-specific graph representations where the weights are determined by the attention scores obtained between each layer-specific representation and a learnable global context vector $\mathbf{c}$. 

\begin{eqnarray}\label{eq:attn_val_gnnlayer}
\psi^{l} = \frac{\exp{(score(\mathbf c, \mathbf g^l))}}{\sum_{j=1}^{L}\exp{(score(\mathbf c,\mathbf g^j))}}
\end{eqnarray}
\begin{eqnarray}
score(\mathbf c, \mathbf g^l) = \frac{{\mathbf c}^\top \mathbf g^l}{\sqrt{d'}}
\end{eqnarray}
\begin{eqnarray}\label{eq:attn_wsum_gnnlayer}
\mathbf z = \sum_{l=1}^L \psi^{l}\mathbf g^l
\end{eqnarray}

where $d'$ is the embedding dimension of the graph's learned representation $\mathbf g^l$ at layer $l$. Therefore, for each \textit{Drug A} and \textit{Drug B} in an input sample triplet, two separate drug representation vectors are computed, $\bz_a$ and $\bz_b$, respectively.

\subsection{Gene Rescaling Model}
Each gene expression vector contains gene expression measurements for $908$ distinct \textit{landmark} genes. We use normalized expression levels of landmark genes to represent the cell line which we refer to as gene expression vector $\mathbf u_e$. To learn a good representation of the cell line, we first embed each gene to a $c$ dimensional space $\mathbf R^c$,  which results in a set of gene vector $\mathbf u_{embed} = \{\mathbf e_1, \mathbf e_2, \cdots, \mathbf e_T\}$ where $T$ is the number of genes and $\mathbf e_i\in \mathbf R^c$.

A global context vector $\mathbf r\in \mathbf R^c$ whose parameters are optimized during training is then applied to each of the gene embedding vectors.  Given the embedding vector $\mathbf e_i$ for 
$i$-th gene,  a score 
is learned by calculating the pairwise similarity between the context vector $\mathbf r$ and the gene embedding vector $\mathbf e_i$. 
Subsequently, these scores are normalized by $\mathrm{Softmax}$ and generate the  $T$ dimensional weight vector $\zeta = [\zeta_1, \zeta_2, \dots, \zeta_T]$ which is used to learn the final representation of the cell line. 

\begin{eqnarray}\label{eq:attn_val_poslayer}
\zeta_{i} = \frac{\exp{(score(\mathbf r, \mathbf e_i))}}{\sum_{j=1}^{T}\exp{(score(\mathbf r,\mathbf e_j))}},
\end{eqnarray}

\begin{eqnarray}
score(\mathbf r, \mathbf e_i) = \frac{{\mathbf r}^\top \mathbf e_i}{\sqrt{l}}
\end{eqnarray}

The final representation of the cell line is given by the weighted sum of embedded gene  vectors:
\begin{eqnarray}\label{eq:attn_wsum_poslayer}
\mathbf z_e = \sum_{t=1}^T \zeta_{t}\mathbf e_t \:\:\:\: \textit{where} \:\:\:\mathbf z_e \in \mathbf R^c
\end{eqnarray}


\subsection{Classifier Model}
The feature representation vectors denoted as $\mathbf{z_a}$ and $\mathbf{z_b}$ for \textit{Drug A} and \textit{Drug B} respectively along with the rescaled gene expression feature vector $\mathbf{z_e}$  are concatenated together to form a unified representation. Subsequently, this concatenated vector is fed into the Classifier model, which consists of a feed-forward neural network (\textit{FNN}) that outputs the log probability distribution over whether a sample (triplet) is more likely to be synergistic or not. 

\subsection{Objective function} 
With a total of $N$ samples $D=\{\bx_i, y_i\}_{i=1}^N$ where $\bx_i$ represents the input triplet (\textit{Drug A}, \textit{Drug B}, \textit{Cell Line}), and $y_i \in \{0,1\}$ represents the class label, we use the negative log-likelihood which is equivalent to binary cross-entropy as the loss function. For each $i$-th sample, the loss is computed based on its true label $y_i$ and the corresponding predicted probability $f_{\theta}(\bx_i)$, representing the probability of the $i$-th sample having a class label $y=1$. This is expressed as:
\begin{equation}
\mathcal{L}(\bx_i, \by_i; \theta) = - [y_i\log f_{\theta}(\bx_i) + (1-y_i)\log(1- f_{\theta}(\bx_i)]\nonumber
\end{equation}
Consequently, the overall loss, denoted as $\mathcal{L}(D; \theta)$, is calculated with an additional $L_2$ regularization term applied to the model parameters, represented as:
\begin{eqnarray}
    \mathcal L(D; \theta) = \frac{1}{N}\sum_{i=1}^N \mathcal L(\bx_i, \by_i; \theta) + \lambda \|\theta\|_{l2}
\end{eqnarray}
where $\lambda$ serves as a hyper-parameter, determining the strength of the regularization effect on the model parameters.

For model training, we utilized a \textit{GNN}-specific format of mini-batching, in which graph adjacency matrices are stacked in a diagonal fashion. This creates a giant graph that contains multiple isolated subgraphs. In our graph-based setting, common mini-batching approaches would be suboptimal, as they would likely result in unnecessary memory consumption. 

\section{Experiment setup and results}
\subsection{Benchmark datasets}
We obtained the drug combination dataset from the \textit{DrugComb}~\citep{zheng2021drugcomb} database\footnote{\tiny{\url{https://drugcomb.org/}}} (version $1.5$). This dataset comprises an initial set of 1,432,351 samples (i.e., \textit{Drug A}, \textit{Drug B}, \textit{Cell Line}) combination triplets. These samples are drawn from 34 distinct studies, including notable sources such as \textit{NCI-ALMANAC}~\citep{holbeck2017national}, \textit{O'NEIL} (\textit{Merck})~\citep{o2016unbiased} and \textit{CLOUD}~\citep{licciardello2017combinatorial}. For each of the samples, we consider the chemical structures of the drugs, the gene expression profiles of untreated cell lines, and four different synergy scores: \emph{Loewe, Bliss, HSA, ZIP} (see Appendix section ~\ref{sec:synergy_score}).

The initial dataset underwent the following steps: Triplets with corresponding identifiers in the gene expression features table were selected. Samples with missing values in drug or cell line identifiers, absent synergy scores, or duplicated triplets were filtered out. Subsequently, synergistic and antagonistic classes (i.e. labels) were defined for each synergy score using thresholding based on the \textit{SynergyFinder} software's documentation\footnote{\tiny{\url{https://synergyfinder.fimm.fi/synergy/synfin_docs/\#datanal}}}, applying recommended thresholds ($\geq 10$ for synergy and $\leq -10$ for antagonism) individually to the four synergy scores.

 Each computed synergy score has inherent biases and limitations due to varied calculation assumptions. For instance, the Loewe synergy score, relying on the \textit{Dose Equivalence Principle} (DEP framework), artificially inflates synergy scores through a Hill-slope dependent bias~\citep{wooten2021musyc}. To address this, we created three majority-voting datasets (Majority-4, Majority-3, Majority-2). Samples are categorized as positive or negative based on whether at least all, three, or two out of the four synergy score measures vote them as such. This approach ensures consistent categorization across thresholded scores, enhancing confidence in identifying synergy or antagonism. The resulting seven datasets, detailed in Table~\ref{table:datasets}, were utilized for both model training and evaluation. Notably, in Majority-2, instances exceed the minimum required for each synergy measure dataset, as inclusion is based on consistent labels from any two synergy measures.

\begin{table}[!ht]
\begin{center}
\resizebox{\linewidth}{!}{
    \begin{tabular}{l|c|c|c|c}
    \hline
        Dataset & $\#$ Samples & Positive Labels & $\#$ Drugs & $\#$ Cell Lines \\ \hline
        Loewe      & $166'608$             & $\approx15\%$                   & $2'148$               & $169$               \\
        Bliss      & $126'619$             & $\approx49\%$                   & $1'869$               & $169$               \\
        HSA      & $109'691$             & $\approx30\%$                   & $1'190$               & $167$               \\
        ZIP      & $89'868$             & $\approx40\%$                  & $1'811$               & $167$               \\
        Majority-4      & $26'031$             & $\approx50\%$                   & $895$               & $153$               \\
        Majority-3      & $64'168$             & $\approx36\%$                   & $987$               & $163$               \\
        Majority-2      & $110'340$             & $\approx40\%$                   & $1'673$               & $167$               \\
        \hline
    \end{tabular}
    }
    \end{center}
    \caption{Benchmark datasets}
    \label{table:datasets}
\end{table}
\paragraph{Drug molecule graph representation}


For each drug in the \textit{DrugComb} database, a chemical representation string ($\mathrm{SMILES}$) was included. Using the \textit{RDKit} cheminformatics software package~\citep{landrum2016rdkit} we then extracted the following features for each atom in the chemical structure of a drug: 
$1)$ Atomic number: Number of protons found in the nucleus of this atom.
$2)$ Chirality (chiral tag): Whether they are superimposable on their mirror image or not.
$3)$ Degree of the atom in the molecule including Hs (hydrogen atoms). The degree of an atom is defined to be its number of directly-bonded neighbors. The degree is independent of bond orders.
$4)$ Formal charge.
$5)$ Total number of Hs (explicit and implicit) on the atom.
$6)$ Number of radical electrons.
$7)$ Hybridization of the atom.
$8)$ Whether the atom is aromatic or not.
$9)$ Whether the atom is in the ring or not.

Therefore, for a given chemical structure of a drug with $N_a$ atoms we constructed a $N_a \times 9$ node feature matrix, where $N_a$ is the number of atoms (nodes) of the chemical structure graph. Additionally, we constructed for each drug an $N_a \times N_a$ binary adjacency matrix $A$, which contained the structural information of each chemical compound. Particularly, this matrix has a value of $1$ at each position where two atoms are forming a bond, otherwise it is $0$s.

\paragraph{Cell line representation}


For this study, we retrieved untreated cell line features from cancerrxgene\footnote{\tiny{\url{https://www.cancerrxgene.org/gdsc1000//GDSC1000_WebResources//Data/preprocessed/Cell_line_RMA_proc_basalExp.txt.zip}}}. This includes the RMA normalized basal expression profiles for all the cell lines included in the raw data by~\cite{iorio2016landscape}, which contains transcriptional profiles of roughly $1'000$ human cancer cell lines ($\mathrm{E-MTAB-}3610$ in \textit{ArrayExpress}). For each cell line, the normalized expression levels of $17'737$ genes were included, out of which we selected $908$ \textit{landmark} genes. The latter were accessed from the $\mathrm{L1000}$~\citep{subramanian2017next} project \footnote{\tiny{\url{https://clue.io/}}}. See Supplementary Files for all included genes. Hence, each cell line is represented by a feature vector of length $908$. See Appendix section \ref{sec:genes} for all included genes.

\begin{table*}[!ht]
    \centering
    \resizebox{\textwidth}{!}{
        \begin{tabular}{l|cccccccc}
        \hline
            \textbf{Model} & \textbf{DDoS} & \textbf{GNN-GAT} & 
            \textbf{DeepDDS}&
            \textbf{DeepSynergy} & \textbf{Logistic R.} & \textbf{Decision T.} & \textbf{KNN} & \textbf{Gradient B.} \\ \hline
            \textbf{ZIP} & \textbf{0.948} & 0.914 & 0.930&0.859 & 0.690 & 0.836 & 0.845 & 0.784 \\
            \textbf{Loewe} & \textbf{0.777} & 0.597 & 0.698&0.544 & 0.371 & 0.567 & 0.524 & 0.510 \\
            \textbf{Bliss} & \textbf{0.930} & 0.886 &0.905& 0.807 & 0.611 & 0.785 & 0.775 & 0.707 \\
            \textbf{HSA} & \textbf{0.921} & 0.865 & 0.891&0.789 & 0.607 & 0.732 & 0.746 & 0.709 \\
            \textbf{Majority-$4$} & \textbf{0.964} & 0.932 &0.933& 0.854 & 0.703 & 0.837 & 0.845 & 0.844 \\
            \textbf{Majority-$3$} & \textbf{0.963} & 0.917 &0.956& 0.825 & 0.593 & 0.792 & 0.806 & 0.751 \\
            \textbf{Majority-$2$} & \textbf{0.953} & 0.903 &0.921& 0.807 & 0.575 & 0.770 & 0.771 & 0.701 \\
            \hline
        \end{tabular}
   }
    \caption{\centering{\small{AUPR scores for different models and datasets.}}}
    \label{table:AUPR_scores}
\vspace{1em}
    \centering
    \resizebox{\textwidth}{!}{
        \begin{tabular}{l|cccccccc}
        \hline
            \textbf{Model} & \textbf{DDoS} & \textbf{GNN-GAT} &             \textbf{DeepDDS}&
            \textbf{DeepSynergy} & \textbf{Logistic R.} & \textbf{Decision T.} & \textbf{KNN} & \textbf{Gradient B.} \\ \hline
            \textbf{ZIP AUC} & \textbf{0.950} & 0.918 &0.932& 0.864 & 0.671 & 0.763 & 0.830 & 0.780 \\
            \textbf{Loewe} & \textbf{0.929} & 0.861 & 0.902&0.836 & 0.748 & 0.718 & 0.789 & 0.817 \\
            \textbf{Bliss } & \textbf{0.951} & 0.915 &0.929& 0.849 & 0.656 & 0.749 & 0.808 & 0.752 \\
            \textbf{HSA} & \textbf{0.956} & 0.921 & 0.936&0.874 & 0.718 & 0.766 & 0.836 & 0.810 \\
            \textbf{Majority-$4$} & \textbf{0.973} & 0.947 &0.948& 0.889 & 0.745 & 0.821 & 0.874 & 0.877 \\
            \textbf{Majority-$3$} & \textbf{0.978} & 0.948 &0.972& 0.889 & 0.697 & 0.800 & 0.866 & 0.830 \\
            \textbf{Majority-$2$} & \textbf{0.968} & 0.933&0.943 & 0.867 & 0.664 & 0.765 & 0.826 & 0.777 \\
            \hline
        \end{tabular}
    }
    \caption{\centering{\small{AUC scores for different models and datasets.}}}
    \label{table:AUC_scores}
\end{table*}

\subsection{Model evaluation}
\textbf{Baseline models}
We compared various baseline models, including classical machine learning and deep learning models. To assess the advantage of the graph-based representation over hand-engineered physicochemical features, we conducted comparisons with \textit{DeepSynergy}~\citep{Preuer2018}. Moreover, we also compared it against the vanilla GNN-based model, which employs the commonly utilized $\mathrm{GAT}$ operator, to test if the introduced attention-based graph network boosts model performance. Finally, we also compared it against DeepDDs \citep{wang2022deepdds}.

\textbf{Model evaluation} We evaluated our proposed model performance on the seven processed datasets as shown in Table \ref{table:datasets} (four given synergy scores and three combined majority voting scores). We used standard binary classification metrics,  \textit{AUC} and \textit{AUPR}. AUC is the Area Under the True Positive Rate-False Positive Rate Curve, while AUPR is the Area Under the Precision-Recall Curve.
Although the majority of datasets we assessed exhibit a balanced distribution (refer to Table~\ref{table:datasets}), it's noteworthy that the \emph{Loewe} dataset comprises only around $15\%$ positive (synergistic) labels. In instances of prevalent class imbalances, the \textit{AUPR} evaluation score is considered to be a more equitable metric. 

In the model training process, we adopted a Stratified 5-Folds cross-validation strategy. This method ensures that the test split maintains a balanced representation of samples for each class, preserving the proportionality of class distributions in each train-test split. Additionally, $10\%$ of the training partition of each fold was reserved for validation and hyperparameter tuning. We initially selected random hyperparameter values for the training of each model on a random fold (out of $5$ folds). Subsequently, we repeated the training of each model on all $5$ folds based on the best-performing hyperparameters of the initial random fold. Finally, the trained models (on all $5$ folds) were tested on the corresponding test split. 


\subsection{Results}
We compared \textit{DDoS} (our model) to six baseline models 
  on seven derived datasets, four of which are based on individual drug synergy scores and three with combined scores. As shown in Tables~\ref{table:AUPR_scores} and ~\ref{table:AUC_scores}, our model outperforms the baseline models. When compared to DeepSynergy (i.e. a deep learning model), our model has better performance on all seven datasets when comparing both AUPR and AUC scores. 
Furthermore, when comparing our model to other classical machine learning baseline models, we observed similar and larger gap (i.e., differences in scores) using both metrics (AUPR and AUC). Lastly, our model still outperforms a GNN-based models (\textit{GNN-GAT}) and DeepDDS on all seven datasets in both AUC and AUPR scores (Table~\ref{table:AUPR_scores} and ~\ref{table:AUC_scores}). The difference in performance ranges from $3$ to $18$ points in AUPR, and from $3$ to $6$ points in AUC scores respectively favoring our model.

For further model evaluation, we generated four additional benchmark datasets, one for each of the four synergy scores (Loewe, Bliss, HSA, ZIP). These datasets differ from the reported benchmark dataset above, by including the additive triplets in their respective non-synergistic class (negative class), i.e., triplets which have synergy values between the two specified thresholds, $-10$ and $10$ (see \textit{Benchmark datasets} section for more details). Those datasets are vastly more imbalanced since the new non-synergistic class includes a large number of additional samples. Therefore, those resulting Loewe, Bliss, HSA and ZIP datasets include $5.3\%$, $13.8\%$, $6.5\%$ and $11.9\%$ positive (synergistic) labels, respectively. This imbalance would be challenging for models' performance when using AUPR metric. Our model still outperforms the DeepSynergy baseline model on those four additional datasets as reported in Table~\ref{table:model_scores_DS_additive}. 
\begin{table*}[!ht]
    \centering
    \resizebox{0.7\textwidth}{!}{
        \begin{tabular}{l|cc|cc|cc|cc}
            \hline
            & \multicolumn{2}{c|}{\textbf{ZIP}} & \multicolumn{2}{c|}{\textbf{Loewe}} & \multicolumn{2}{c|}{\textbf{Bliss}} & \multicolumn{2}{c}{\textbf{HSA}} \\ \hline
            Model & AUPR & AUC & AUPR & AUC& AUPR & AUC & AUPR & AUC \\ \hline
            \textbf{DDoS} &\textbf{ 0.795} & \textbf{0.951} & \textbf{0.406} & \textbf{0.900} & \textbf{0.824} & \textbf{0.948} & \textbf{0.522 }& \textbf{0.915} \\
            \textbf{DeepSynergy} & 0.574 & 0.891 & 0.235 & 0.835 & 0.587 & 0.878 & 0.297 & 0.829 \\
            \hline
        \end{tabular}
    }
    \caption{Model evaluation scores for additional datasets with additive samples.}
    \label{table:model_scores_DS_additive}
\end{table*}

\paragraph{Different train/test split}
Conventional train-test data splits may fall short of an ideal scenario, merely ensuring the exclusion of triplets (drug A, drug B, and cell line) observed in the training set from the test set. However, they do not guarantee the absence of certain drugs or cell lines in the training set. To address this limitation, we expanded our model training to incorporate various data split strategies. Specifically, we excluded all triplets containing certain cell lines or drugs from the training set. This meticulous approach ensures that the training data is devoid of any triplets involving the cell lines/drugs on which the model is being trained. As table \ref{table:model_scores_new_split} shows, our model demonstrates its ability to predict on the cell lines or drugs that are fully unseen during the training process. This indicates the robustness of our model in handling scenarios with a more controlled and stringent data split, providing a more reliable evaluation of its predictive capabilities.

\begin{table*}[!ht]
    \centering
    \resizebox{0.9\textwidth}{!}{
        \begin{tabular}{l|cc|cc|cc|cc}
            \hline
            & \multicolumn{2}{c|}{\textbf{ZIP}} & \multicolumn{2}{c|}{\textbf{Loewe}} & \multicolumn{2}{c|}{\textbf{Bliss}} & \multicolumn{2}{c}{\textbf{HSA}} \\ \hline
         \textbf{Split Strategy} & \textbf{AUPR} & \textbf{AUC} & \textbf{AUPR} & \textbf{AUC} & \textbf{AUPR} & \textbf{AUC} & \textbf{AUPR} & \textbf{AUC}  \\ \hline
            Cell Lines & 0.764 $\pm$0.052 & 0.713$\pm$0.045& 0.508$\pm$0.027 & 0.831 $\pm$0.027& 0.652$\pm$ 0.063 & 0.649$\pm$ 0.041 & 0.697$\pm$0.057 & 0.799$\pm$0.025 \\
            \hline
            Drug 1 & 0.886$\pm$0.038& 0.843$\pm$0.001 & 0.460$\pm$0.023 & 0.822$\pm$0.001 & 0.807$\pm$0.026 & 0.815 $\pm$ 0.034& 0.725$\pm$0.039 & 0.855 $\pm$0.012\\
            \hline
            Drug 2 & 0.894$\pm$ 0.020 & 0.855$\pm$0.001& 0.507$\pm$0.029 & 0.822 $\pm$0.020& 0.805$\pm$0.025 & 0.807 $\pm$ 0.014& 0.714 $\pm$ 0.015& 0.855 $\pm$0.014\\
            \hline
        \end{tabular}
    }
    \caption{Performance evaluation of our DDoS model across various train-test split strategies: the `Cell Lines' row represents the model's performance when specific cell lines are excluded from the training set and used for testing. Similarly, the ``Drug 1'' row signifies the model's evaluation when certain drugs from the Drug 1 set are omitted during training.}
    \label{table:model_scores_new_split}
\end{table*}

\subsection{Case Studies}
To further validate our model, we examined the top $20$ novel synergistic predictions of drug-drug-cell line combinations (from the test set), ranked by the highest synergy probability from our \textit{DDoS} model. For consistency with previous studies, we considered the top ``false positives'' of the Loewe dataset and investigated how strongly these predictions are supported by the remaining synergy scores, i.e., Bliss, ZIP and HSA. In Table~\ref{table:case_studies} we list those triplets along with an indication of how many of the three scores have a positive score for each triplet. We discovered that $75\%$ of the top triplets have at least one other synergy score (Bliss, ZIP, HSA) supporting those predictions (i.e. confirming synergism) while in nearly $50\%$ of the cases, all three scores have this indication, highlighting the efficacy of our model by representing all synergy scores (i.e. less biased toward one vs. the other).

\begin{table*}[!ht]
\centering
\small
\begin{tabular}{llllr}
\hline
 Rank & Drug A & Drug B & Cell Line & Indication \\
\hline
1 & Erlotinib & 7-ethyl-10-hydroxycamptothecin & NCIH520 & *** \\
2 & Dimesna & Temozolomide & T98G & - \\
3 & Crizotinib & Actinomycin d & SK-MEL-2 & - \\
4 & Uramustine & Actinomycin d & M14 & - \\
5 & 391210-10-9 & Deforolimus & OCUBM & *** \\
6 & Crizotinib & Mithramycin & SK-MEL-2 & *** \\
7 & Sunitinib & 891494-63-6 & NCIH2122 & *** \\
8 & Geldanamycin & 915019-65-7 & OCUBM & *** \\
9 & Foscarnet sodium & Temozolomide & T98G & - \\
10 & Vemurafenib & Mithramycin & RPMI-8226 & *** \\
11 & Paclitaxel & Lapatinib & RPMI7951 & *** \\
12 & Actinomycin d & Antibiotic ad 32 & M14 & *** \\
13 & Geldanamycin & Topotecan & A375 & - \\
14 & Crenolanib & Ncgc00162423-03 & L-1236 & ** \\
15 & Antibiotic ay 22989 & Adm hydrochloride & KM12 & ** \\
16 & Docetaxel & Lenalidomide & RVH-421 & ** \\
17 & Docetaxel & Zm 336372 & WM115 & *** \\
18 & 1032350-13-2 & Nilutamide & SK-MEL-28 & ** \\
19 & Cyclophosphamide & Temozolomide & T-47D & ** \\
20 & Vandetanib & Mithramycin & HL-60(TB) & * \\
\hline
\end{tabular}
\caption{\centering{Case studies - Top 20 ``false positives'' predictions of the Loewe dataset, ranked by the model probability of synergistic outcome. The number of $*$ indicates how many of the HSA, ZIP, and Bliss scores support each prediction.}}
\label{table:case_studies}
\end{table*}

\section{Calibration analysis }
\begin{figure}[H]
    \centering
    \includegraphics[scale=0.4]{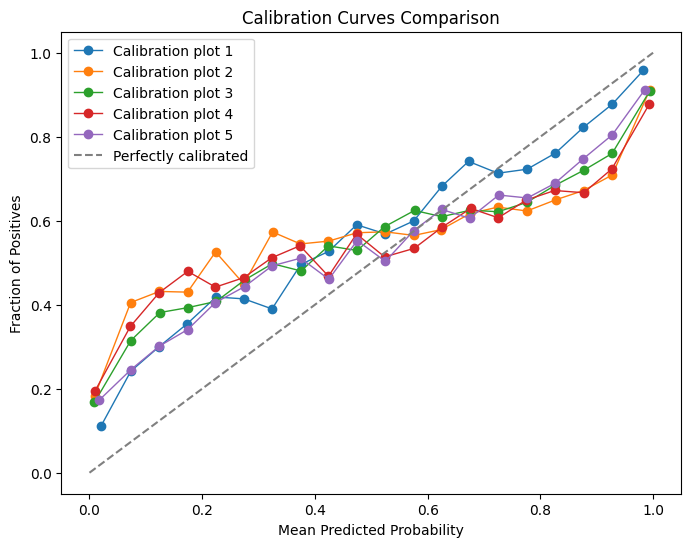}
    \caption{Calibration Analysis }
    \label{fig:calibiration}
\end{figure}

In our study, we also conducted a calibration analysis to compare the predicted probabilities of synergy against the observed frequencies of actual synergistic outcomes. Figure \ref{fig:calibiration} showcases the calibration curve derived from the results of our trained model, specifically for the ZIP synergy score. This model was trained using a leave-some-drugs-out approach, as detailed in Table \ref{table:model_scores_new_split}. The figure illustrates five distinct calibration plots, each corresponding to one of the five models obtained through 5-fold cross-validation. These plots serve to visually assess the accuracy of our probabilistic predictions in relation to the true synergistic interactions observed. Please note that this analysis corresponds to the model that exhibited relatively lower performance, as indicated in Table \ref{table:model_scores_new_split}. This outcome is primarily attributed to our ``leave-drugs-out'' training and testing strategy, wherein certain drugs were omitted from the training set, allowing the model to be evaluated on previously unseen drugs. Consequently, for the other models, which are detailed in Table \ref{table:AUC_scores}, we anticipate a more robust calibration performance, as they were not subjected to this specific train-test data split approach.

\section{Conclusion}
Our drug synergy prediction model, \textit{DDoS}, employs an \textit{end-to-end} architecture to learn task-specific representations of drugs and cell lines. In contrast, the prevailing approaches rely on task-agnostic, pre-computed chemical characteristics of drugs \citep{Preuer2018}, \citep{zhang2021synergistic}, \citep{Kuru2020}, coupled with suboptimal representations of gene expression features. By leveraging a \textit{GNN}-based model, we extract task-specific drug chemical representations, while gene expression vectors are dynamically rescaled during model training. Our approach \textit{DDoS} has demonstrated clear advantages, as evidenced by the improved performance observed in our reported experiments.

For model training, we utilized data from a large drug combination database which contains nearly $1.5M$ drug-drug-cell line combination triplets and four distinct drug synergy scores. Conversely, the baseline model, DeepSynergy, was originally trained only on $23'062$ samples~\cite{Preuer2018}. Thus, our model was able to learn from the most complete drug combination dataset to date, including thousands of drugs and hundreds of cell lines.  Moreover, due to the different underlying assumptions of drug synergy scores, each score includes its distinct biases and limitations. This leads to multiple different labels (synergistic or not) assigned to the same samples, depending on which synergy score is utilized. Thus, only a fraction of samples are characterized as synergistic by all four scores simultaneously (see Majority-$4$ dataset in Table~\ref{table:datasets}). By combining these scores through majority voting, we constructed and trained on three high-confidence datasets in addition to the four individual drug synergy scores.

Despite the multiple advantages, our solution also encompasses some limitations. The cell line features we utilized are only based on gene expression data. For a more informative representation, further cell line characteristics can be included, e.g., mutations, copy number variations, pathways, drug targets, and more. Additionally, majority voting-based combinations of synergy scores alleviate some of the biases of each score. Ideally, improved single-drug synergy scores would provide these advantages. 
Furthermore, the binary classification task (synergistic vs. non-synergistic) could be expanded to include further classes and make predictions on the efficacy and potential toxic effects of drug combinations. Moreover, since the dataset predominantly comprises drugs and cell lines related to cancer treatments, conducting experiments for other complex diseases (e.g., viral infections) would mitigate the cancer-specific bias in our model.
\paragraph{Future outlook}
Detecting synergistic effects among approved drugs holds clinical significance, as drug repurposing can streamline the expensive and lengthy process of developing new drugs. Often, novel drugs may not demonstrate efficacy in clinical trials despite thorough preclinical safety testing. In the future, integrating in silico drug synergy discovery with \textit{ex vivo} testing of selected anti-cancer drug combinations in patient-derived $\mathrm{3D}$ organoid models could significantly enhance the clinical translatability of this approach.

\acks{This work is supported by the Swiss National Science Foundation (project 201184) and  Swiss 3R Competence Center Doctoral Grant. 
}

\newpage
\bibliography{chil-sample}

\newpage
\appendix

\section{Database}\label{sec:database}
Notable examples include the \textit{NCI-ALMANAC} dataset~\citep{holbeck2017national} which contains $103$ FDA-approved drugs tested in $60$ different cell lines (NCI-$60$)~\citep{Holbeck2017} and the large oncology dataset produced by \textit{Merck $\&$Co}~\citep{ONeil2016} which is composed of $38$ drugs tested in $39$ different cell lines from $6$ different tissue types. Furthermore, databases like \textit{DrugComb}~\citep{zheng2021drugcomb} have collected and curated drug combination synergy data from $34$ different sources, thus comprising $8'397$ different drugs, $2'320$ cell lines, and $739'964$ drug combinations in total.
\section{Drug synergy reference scores}\label{sec:synergy_score}
Drug synergy refers to a quantification of drug interaction~\citep{geary2013understanding}, i.e., how much excess drug response (e.g., tumor cell death) is observed compared to the expectation (that is, no excess). Thus, the reference synergy scores (Loewe, Bliss, HSA, ZIP) provide a percentage of excess response, or reduced response in the antagonistic setting.

\paragraph{Loewe Additivity Model}
The Loewe additivity model is based on the concepts of sham combination and dose equivalence. Therefore, the expected effect ${y_{Loewe}}$ can be defined as if a drug was combined with itself, i.e., ${y_{Loewe} = y_{1}(x_{1}+x_{2}) = y_{2}(x_{1}+x_{2})}$, where ${y_1}$ and ${y_2}$ indicate the drug responses of drug $1$ and drug $2$, respectively. This model considers the dose-response curves of individual drugs, where the expected effect must satisfy:
\begin{equation}
  \frac{x_{1}}{x_{Loewe}^{1}} + \frac{x_{2}}{x_{Loewe}^{2}} = 1,
\label{eq:loewe}
\end{equation}

where ${x_{1}}$, ${x_{2}}$ are drug doses and ${x_{Loewe}^{1}}$ ,${x_{Loewe}^{2}}$ are the doses of drug $1$ and $2$ alone that produce ${y_{Loewe}}$. By using $4$-parameter log-logistic ($\mathrm{4PL}$) curves to describe dose-response curves the following parametric form of previous equation is derived:

\begin{equation}
  \frac{x_1}{m_1(\frac{y_{Loewe}-E_{\min}^{1}}{E_{\max}^{1} - y_{Loewe}})^{\frac{1}{\lambda_1}}} + \frac{x_2}{m_2(\frac{y_{Loewe}-E_{\min}^{2}}{E_{\max}^{2} - y_{Loewe}})^{\frac{1}{\lambda_2}}} = 1,
\label{eq:loewe_derived}
\end{equation}

where $E_{\min}, E_{\max} \in [0,1]$ are minimal and maximal effects of the drug, $m_{1}$, $m_{2}$ are the doses of the drug that produce the midpoint effect of  $E_{\min} +  E_{\max}$, also known as relative $EC_{50}$ or  $IC_{50}$, and $\lambda_1(\lambda_1 >0)$, $\lambda_2(\lambda_2 >0)$ are the shape parameters indicating the sigmoidicity or slope of dose-response curves. A numerical nonlinear solver can be then used to determine $y_{Loewe}$ for ($x_1$, $x_2$).

\paragraph{Bliss independence model}
The Bliss independence model is based on a stochastic process. It assumes that each drug has different response in a particular scenario and that they elicit their effects independently. Therefore, the combination response is based on the probability of those independent events. Hence, the Bliss score formula is as follows:
\begin{equation}
y_{Bliss} = y_1 + y_2 - y_1 y_2,
\label{eq:bliss}
\end{equation}

where ${y_{Bliss}}$ represents the Bliss response. ${y_1}$ and ${y_2}$ indicate the drug responses of drug $1$ (${D_1}$) and drug $2$ (${D_2}$), respectively.

\paragraph{Highest Single Agent Model}
The highest single agent model (HSA) states that the effect of the expected combination effect is equal to the maximum effect of each individual drug at specific concentrations. Thus, the formula representing the model is:

\begin{equation}
    y_{HSA} = max(y_1, y_2),
\label{eq:hsa}
\end{equation}
where ${y_1}$ and ${y_2}$ are the monotherapy (single drug) effect of drug 1 (${D_1}$) and drug 2 (${D_2})$, respectively.

\paragraph{Zero Interaction Potency (ZIP)}
The Zero Interaction Potency calculates the expected effect of two drugs under the assumption that they do not potentiate each other, i.e., both assumptions of the Loewe model and the Bliss model are met:

\begin{equation}
y_{ZIP} = \frac{(\frac{x_1}{m_1})^{\lambda_{1}}}{1+(\frac{x_1}{m_1})^{\lambda_{1}}} +  \frac{(\frac{x_2}{m_2})^{\lambda_{2}}}{1+(\frac{x_2}{m_2})^{\lambda_{2}}} - \frac{(\frac{x_1}{m_1})^{\lambda_{1}}}{1+(\frac{x_1}{m_1})^{\lambda_{1}}} \frac{(\frac{x_2}{m_2})^{\lambda_{2}}}{1+(\frac{x_2}{m_2})^{\lambda_{2}}}
\label{eq:ZIP}
\end{equation}

where ${x_1}$, ${x_2}$, ${m_1}$, ${m_2}$, ${\lambda_1}$ and ${\lambda_2}$, are defined in the Loewe model section.

\subsection{Optimization}
The hyperparameters utilized for the training of all models are, Batch size: $300$, Epochs: $100$, Graph embedding dimension: $100$, Number of \textit{GNN} layers ($L$): $5$, Dropout: $0.3$, Learning rate: $0.0003$, Weight decay: $0.0$, Hidden layer $1$ size: $4096$, Hidden layer $2$ size: $1024$. 

\section{Hyper parameter tuning}
We employed random search to tune the hyperparameters, constrained by our computational resource limits. To manage these constraints effectively, we fixed the majority of the hyperparameters to predetermined values. However, we selectively varied a few key hyperparameters within specific ranges. For example, one such hyperparameter was set to explore values within the set \{512, 1024, 2048, 4096\} as it is presented in Table \ref{table:model_hyperparameters}.
\begin{table}
\centering
\caption{Model Hyperparameters}
\begin{small}
\begin{tabular}{|l|l|}
\hline
\multicolumn{2}{|c|}{\textbf{Training Parameters}} \\
\hline
Batch Size & 300 \\
Number of Epochs & 50 \\
Base Learning Rate & $3 \times 10^{-4}$ \\
Max Learning Rate Multiplier & 5 \\
L2 Regularization & $1 \times 10^{-5}$ \\
Loss Weight & 1.0 \\
Margin Value & 1.0 \\
\hline
\multicolumn{2}{|c|}{\textbf{Network Architecture}} \\
\hline
Embedding Dimension & 300 \\
GNN Type & ``gatv2" \\
Number of Layers & 5 \\
Graph Pooling & \{``mean", ``max"\} \\
Input Embedding Dimension & None \\
Gene Embedding Dimension & 1 \\
Number of Attention Heads & 2 \\
Number of Transformer Units & 1 \\
Dropout Probability & 0.3 \\
MLP Embedding Factor & 2 \\
Pooling Mode & ``attn" \\
Distance Option & ``cosine" \\
\hline
\multicolumn{2}{|c|}{\textbf{Expression Model Parameters}} \\
\hline
Expression Dimension & 64 \\
Expression Input Size & 908 \\
Hidden Layer 1 Size & \{$2^9$, $2^{10}$, $2^{11}$, $2^{12}$\} \\
Hidden Layer 2 Size & \{$2^9$, $2^{10}$\} \\
\hline
\end{tabular}
\end{small}
\label{table:model_hyperparameters}
\end{table}




\section{Explainability}
To gain a better understanding of which genes in the cell line representation are more relevant for the prediction of drug synergy, we also implemented SHAP analysis \citep{lundberg2017unified}. For this experiment, to keep the model simple and focus on the cell line representations, we used one of our baselines, DeepSynergy,  which is a feed-forward neural network. We modified it so that for the cell line representation it uses the the expression values of 908 landmark genes as we did for our model.  We trained it using  Majority-4 datasets and calculated the SHAP values. 

\begin{figure*}
    \centering
    \begin{minipage}{0.49\textwidth}
        \centering
        \includegraphics[scale=0.2]{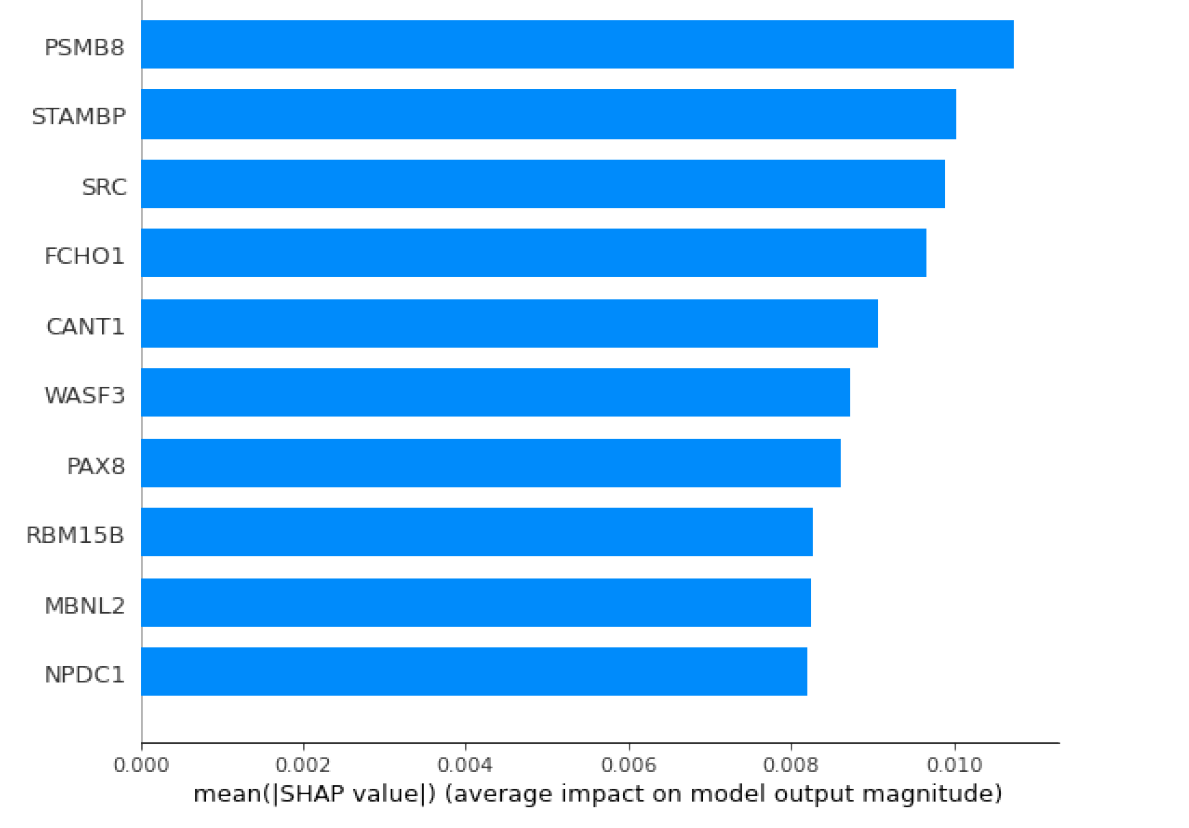}
        \caption{SHAP analysis for gene features}
        \label{fig:shap_gene}
    \end{minipage}
    \hfill 
    \begin{minipage}{0.49\textwidth}
        \centering
        \includegraphics[scale=0.22]{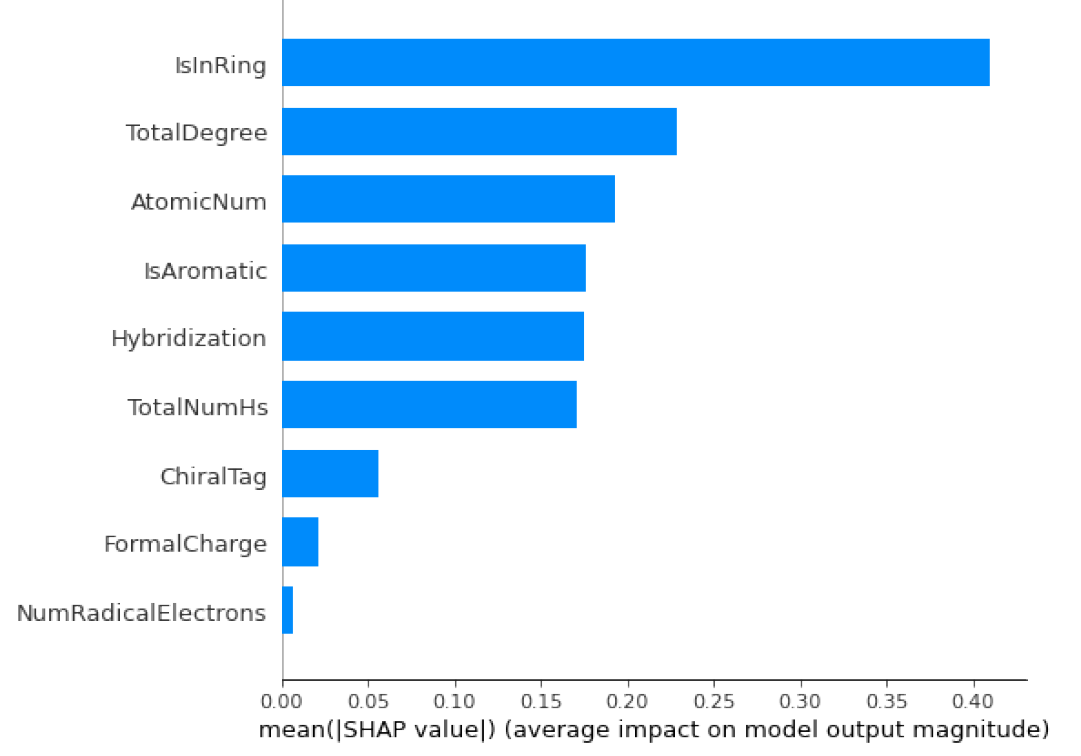}
        \caption{SHAP analysis for drug features}
        \label{fig:shap_drug}
    \end{minipage}
    \caption{SHAP Analysis over the cell line features}
    \label{fig:shap}
\end{figure*}

The input features for which we measured their impact on the model performance are
nine atomic features of the drug chemical structures (for two drugs in each sample) and
the expression values of 908 landmark genes for the cell line of the sample. Thus, the total number of features is equal to 926.
We utilized the GradientSHAP implementation from the Captum library \citep{kokhlikyan2020captum} for the computation of (approximate) SHAP values and convergence deltas.
The top 10 important genes, according to their mean absolute SHAP values, are
illustrated in Figures \ref{fig:shap_gene} and \ref{fig:shap_drug}. Their impact on the model output has a lower range (0.008
- 0.011) compared to the chemical features range (0.01 - 0.40). However, the total
(i.e., summed, as SHAP values are additive) impact of cell line features is larger than
the total impact of drug features (5.486 vs. 1.782, respectively). Consequently, as
the number of cell line features is much larger than the number of drug features (908
vs. 9, respectively), the individual feature impacts are smaller, but the total impact is
larger for the cell lines. Nevertheless, specific drug features have a considerable impact by
themselves, thus indicating which characteristics of drugs could potentially be prioritized
for the selection of synergistic drug combinations. 

\section{List of landmark genes}\label{sec:genes}
{\tiny AARS, ABCC5, ABCF1, ABHD4, ABHD6, ABL1, ACAA1, ACAT2, ACBD3, ACD, ACLY, ACOT9, ADAM10, ADAT1, ADH5, ADI1, ADO, ADRB2, AGL, AKAP8, AKAP8L, AKR7A2, AKT1, ALAS1, ALDH7A1, ALDOA, AMDHD2, ANKRD10, ANO10, ANXA7, APBB2, APOE, APP, APPBP2, ARFIP2, ARHGAP1, ARHGEF12, ARHGEF2, ARID4B, ARID5B, ARL4C, ARNT2, ARPP19, ASAH1, ASCC3, ATF1, ATF5, ATF6, ATG3, ATMIN, ATP11B, ATP1B1, ATP2C1, ATP6V0B, ATP6V1D, AURKA, AURKB, AXIN1, BACE2, BAD, BAG3, BAMBI, BAX, BCL2, BCL7B, BDH1, BECN1, BHLHE40, BID, BIRC2, BIRC5, BLCAP, BLMH, BLVRA, BMP4, BNIP3, BNIP3L, BPHL, BRCA1, BTK, BZW2, C2CD2, C2CD2L, C2CD5, C5, CAB39, CALM3, CALU, CAMSAP2, CANT1, CAPN1, CASC3, CASK, CASP10, CASP2, CASP3, CASP7, CAST, CAT, CBLB, CBR1, CBR3, CCDC85B, CCDC86, CCDC92, CCL2, CCNA1, CCNA2, CCNB1, CCND1, CCND3, CCNE2, CCNF, CCNH, CCP110, CD320, CD40, CD44, CDC20, CDC25A, CDC25B, CDC42, CDC45, CDCA4, CDH3, CDK1, CDK19, CDK2, CDK4, CDK5R1, CDK6, CDK7, CDKN1B, CDKN2A, CEBPD, CEBPZ, CENPE, CEP57, CERK, CETN3, CFLAR, CGRRF1, CHAC1, CHEK1, CHEK2, CHIC2, CHMP6, CHN1, CHP1, CIAPIN1, CIRBP, CISD1, CLIC4, CLPX, CLSTN1, CLTB, CLTC, CNDP2, CNOT4, CNPY3, COASY, COG2, COG4, COG7, COL1A1, COL4A1, COPB2, COPS7A, CORO1A, CPNE3, CPSF4, CREB1, CREG1, CRELD2, CRK, CRKL, CRTAP, CRYZ, CSK, CSNK1A1, CSNK1E, CSNK2A2, CSRP1, CTNNAL1, CTNND1, CTTN, CXCL2, CXCR4, CYB561, CYTH1, DAG1, DAXX, DCK, DCTD, DCUN1D4, DDB2, DDIT4, DDR1, DDX10, DDX42, DECR1, DENND2D, DERA, DFFA, DFFB, DHDDS, DHRS7, DHX29, DLD, DMTF1, DNAJA3, DNAJB1, DNAJB2, DNAJB6, DNAJC15, DNM1, DNM1L, DNMT1, DNMT3A, DNTTIP2, DPH2, DRAP1, DSG2, DUSP11, DUSP14, DUSP22, DUSP3, DUSP4, DUSP6, DYNLT3, DYRK3, E2F2, EAPP, EBNA1BP2, EBP, ECD, ECH1, EDEM1, EDN1, EED, EFCAB14, EGF, EGFR, EGR1, EIF4EBP1, EIF5, ELAC2, ELAVL1, ELOVL6, EML3, ENOPH1, ENOSF1, EPB41L2, EPHA3, EPHB2, EPN2, EPRS, ERBB2, ERBB3, ETFB, ETS1, ETV1, EVL, EXOSC4, EXT1, EZH2, FAH, FAIM, FAM20B, FAM57A, FAM69A, FAS, FAT1, FBXL12, FBXO11, FBXO21, FBXO7, FCHO1, FDFT1, FEZ2, FGFR2, FGFR4, FHL2, FIS1, FKBP14, FKBP4, FOS, FOSL1, FOXJ3, FOXO3, FOXO4, FPGS, FRS2, FSD1, FUT1, FYN, FZD1, FZD7, G3BP1, GAA, GABPB1, GADD45A, GADD45B, GALE, GAPDH, GATA2, GATA3, GDPD5, GFOD1, GFPT1, GHR, GLI2, GLOD4, GLRX, GMNN, GNA11, GNA15, GNAI1, GNAI2, GNAS, GNB5, GNPDA1, GOLT1B, GPATCH8, GPC1, GRB10, GRB7, GRN, GRWD1, GSTZ1, GTF2A2, GTF2E2, GTPBP8, H2AFV, HADH, HAT1, HDAC2, HDAC6, HEATR1, HEBP1, HERC6, HERPUD1, HES1, HIST1H2BK, HIST2H2BE, HK1, HLA-DRA, HMG20B, HMGA2, HMGCR, HMGCS1, HMOX1, HN1L, HOMER2, HOOK2, HPRT1, HS2ST1, HSD17B10, HSPA1A, HSPA4, HSPA8, HSPD1, HTATSF1, HTRA1, HYOU1, IARS2, ICAM1, ICAM3, ICMT, ID2, IDE, IER3, IFNAR1, IFRD2, IGF1R, IGF2BP2, IGF2R, IGFBP3, IGHMBP2, IKBKAP, IKBKB, IKBKE, IKZF1, IL13RA1, IL1B, IL4R, ILK, INPP1, INPP4B, INSIG1, INTS3, IPO13, IQGAP1, ISOC1, ITFG1, ITGAE, ITGB1BP1, ITGB5, JMJD6, JUN, KAT6A, KAT6B, KCNK1, KCTD5, KDM3A, KDM5A, KDM5B, KEAP1, KIAA0100, KIAA0355, KIAA0753, KIAA0907, KIF14, KIF20A, KIF2C, KIF5C, KIT, KLHDC2, KLHL21, KLHL9, KTN1, LAGE3, LAMA3, LAP3, LBR, LGALS8, LGMN, LIG1, LIPA, LOXL1, LPAR2, LPGAT1, LRP10, LRPAP1, LRRC41, LSM5, LSM6, LSR, LYN, LYRM1, MACF1, MALT1, MAMLD1, MAN2B1, MAP2K5, MAP3K4, MAP4K4, MAP7, MAPK13, MAPK1IP1L, MAPK9, MAPKAPK2, MAPKAPK3, MAPKAPK5, MAST2, MAT2A, MBNL1, MBNL2, MBOAT7, MBTPS1, MCM3, MCOLN1, ME2, MEF2C, MELK, MEST, MFSD10, MICALL1, MIF, MKNK1, MLEC, MLLT11, MMP1, MMP2, MOK, MPC2, MPZL1, MRPL19, MRPS16, MRPS2, MSH6, MSRA, MTA1, MTF2, MTFR1, MTHFD2, MUC1, MVP, MYBL2, MYC, MYCBP, MYCBP2, MYL9, MYLK, MYO10, NARFL, NCAPD2, NCK2, NCOA3, NENF, NET1, NFATC3, NFATC4, NFE2L2, NFIL3, NFKB2, NFKBIA, NFKBIB, NFKBIE, NIPSNAP1, NISCH, NIT1, NMT1, NNT, NOL3, NOLC1, NOS3, NOSIP, NOTCH1, NPC1, NPDC1, NPEPL1, NPRL2, NR1H2, NR2F6, NR3C1, NRAS, NRIP1, NSDHL, NT5DC2, NUCB2, NUDCD3, NUDT9, NUP133, NUP85, NUP88, NUP93, NUSAP1, NVL, ORC1, OXA1L, OXCT1, OXSR1, P4HA2, P4HTM, PACSIN3, PAF1, PAFAH1B1, PAFAH1B3, PAICS, PAK1, PAK4, PAK6, PAN2, PAPD7, PARP1, PARP2, PAX8, PCBD1, PCCB, PCK2, PCM1, PCMT1, PCNA, PDGFA, PDHX, PDIA5, PDLIM1, PDS5A, PECR, PEX11A, PFKL, PGM1, PGRMC1, PHGDH, PHKA1, PHKB, PHKG2, PIGB, PIH1D1, PIK3C2B, PIK3C3, PIK3CA, PIK3R3, PIK3R4, PIN1, PIP4K2B, PKIG, PLA2G15, PLA2G4A, PLCB3, PLEKHJ1, PLEKHM1, PLK1, PLOD3, PLP2, PLS1, PLSCR1, PMAIP1, PMM2, PNKP, PNP, POLB, POLE2, POLG2, POLR1C, POLR2I, POLR2K, POP4, PPARD, PPARG, PPIC, PPIE, PPOX, PPP1R13B, PPP2R3C, PPP2R5A, PPP2R5E, PRAF2, PRCP, PRKACA, PRKAG2, PRKCD, PRKCQ, PRKX, PROS1, PRPF4, PRR15L, PRR7, PRSS23, PSIP1, PSMB8, PSMD10, PSMD4, PSME1, PSMF1, PSMG1, PSRC1, PTGS2, PTK2, PTK2B, PTPN1, PTPN12, PTPN6, PTPRC, PTPRF, PTPRK, PUF60, PWP1, PXN, PYCR1, PYGL, RAB11FIP2, RAB21, RAB27A, RAB31, RAB4A, RAD51C, RAD9A, RAE1, RAI14, RALA, RALB, RALGDS, RAP1GAP, RASA1, RB1, RBKS, RBM15B, RBM6, REEP5, RELB, RFC2, RFC5, RFNG, RFX5, RGS2, RHEB, RHOA, RNF167, RNH1, RNMT, RNPS1, RPA1, RPA2, RPA3, RPIA, RPL39L, RPN1, RPS5, RPS6, RPS6KA1, RRAGA, RRP12, RRP1B, RRP8, RRS1, RSU1, RTN2, RUVBL1, S100A13, S100A4, SACM1L, SCAND1, SCARB1, SCCPDH, SCP2, SCRN1, SCYL3, SDHB, SENP6, SERPINE1, SESN1, SFN, SGCB, SH3BP5, SHC1, SIRT3, SKIV2L, SKP1, SLC11A2, SLC1A4, SLC25A13, SLC25A14, SLC25A4, SLC25A46, SLC27A3, SLC2A6, SLC35A1, SLC35A3, SLC35B1, SLC35F2, SLC37A4, SLC5A6, SMAD3, SMARCA4, SMARCC1, SMARCD2, SMC1A, SMC3, SMC4, SMNDC1, SNAP25, SNCA, SNX11, SNX13, SNX6, SNX7, SOCS2, SORBS3, SOX4, SPAG4, SPAG7, SPDEF, SPEN, SPP1, SPR, SPRED2, SPTAN1, SPTLC2, SQRDL, SQSTM1, SRC, SSBP2, ST3GAL5, ST6GALNAC2, ST7, STAMBP, STAP2, STAT1, STAT3, STAT5B, STK10, STK25, STMN1, STX1A, STX4, STXBP1, STXBP2, SUPV3L1, SUV39H1, SUZ12, SYK, SYNE2, SYNGR3, SYPL1, TARBP1, TATDN2, TBC1D9B, TBP, TBPL1, TBX2, TBXA2R, TCEA2, TCEAL4, TCERG1, TCFL5, TCTA, TCTN1, TERF2IP, TERT, TES, TESK1, TEX10, TFAP2A, TFDP1, TGFB3, TGFBR2, THAP11, TIAM1, TICAM1, TIMELESS, TIMM17B, TIMM22, TIMM9, TIMP2, TIPARP, TJP1, TLE1, TLK2, TLR4, TM9SF2, TM9SF3, TMCO1, TMED10, TMEM109, TMEM2, TMEM5, TMEM50A, TMEM97, TNFRSF21, TNIP1, TOMM34, TOP2A, TOPBP1, TOR1A, TP53, TP53BP1, TP53BP2, TPD52L2, TPM1, TRAK2, TRAM2, TRAP1, TRAPPC3, TRAPPC6A, TRIB1, TRIB3, TRIM13, TRIM2, TSC22D3, TSEN2, TSKU, TSPAN3, TSPAN4, TSPAN6, TSTA3, TUBB6, TXLNA, TXNDC9, TXNL4B, TXNRD1, UBE2A, UBE2C, UBE2J1, UBE2L6, UBE3B, UBE3C, UBQLN2, UBR7, UFM1, UGDH, USP1, USP14, USP22, USP6NL, USP7, UTP14A, VAPB, VAT1, VAV3, VGLL4, VPS28, VPS72, WASF3, WDR61, WDR7, WDTC1, WFS1, WIPF2, WRB, XBP1, XPNPEP1, XPO7, YKT6, YME1L1, YTHDF1, ZDHHC6, ZFP36, ZMIZ1, ZMYM2, ZNF131, ZNF274, ZNF318, ZNF395, ZNF451, ZNF586, ZNF589}
\end{document}